\documentclass[twocolumn,showpacs,preprintnumbers,amsmath,amssymb]{revtex4}

\usepackage{graphicx}% Include figure files
\usepackage{epstopdf}
\usepackage{xcolor}
\usepackage{amsmath}
\usepackage{amstext,bm}
\usepackage{dcolumn}% Align table columns on decimal point
\usepackage{bm}% bold math

\newcommand\mse{National Laboratory of Solid State Microstructures and Department of Materials Science
                and Engineering, Nanjing University, Nanjing 210093, China }
\newcommand\cas{Beijing National Laboratory for Condensed Matter Physics,
                and Institute of Physics, Chinese Academy of Sciences, Beijing 100190, China}
\newcommand\shaoxing{ Department of Physics, Shaoxing University, Shaoxing 312000, China}
\newcommand\phy{National Laboratory of Solid State Microstructures and Department of Physics,
                Nanjing University, Nanjing 210093, China}
\newcommand\twa{Department of Physics, National Taiwan University, Taipei 10617, Taiwan}
\newcommand\twb{Physics Division, National Center for Theoretical Sciences, Hsinchu 30013, Taiwan }

\begin{document}

\title{ Quantum topological Hall effect and noncoplanar antiferromagnetism in K$_{0.5}$RhO$_2$ }
\author{Jian Zhou$^{1}$}
\author{Qi-Feng Liang$^2$}
\author{Hongming Weng$^{3}$}
\author{Y. B. Chen$^4$}
\author{Shu-Hua Yao$^1$}
\author{Yan-Feng Chen$^1$} \thanks{Corresponding author: yfchen@nju.edu.cn}
\author{Jinming Dong$^4$}
\author{Guang-Yu Guo$^{5,6}$} \thanks{Corresponding author: gyguo@phys.ntu.edu.tw}

\affiliation{
$^1$ \mse \\
$^2$ \shaoxing\\
$^3$ \cas\ \\
$^4$ \phy \\
$^5$ \twa \\
$^6$ \twb
}

\date{\today}


\begin{abstract}
Quantum anomalous Hall (QAH) phase is a two-dimensional bulk ferromagnetic insulator with a nonzero
Chern number in presence of spin-orbit coupling (SOC) but absence of applied magnetic fields. Associated
metallic chiral edge states host dissipationless current transport in electronic devices.
This intriguing QAH phase has recently been observed in magnetic impurity-doped topological insulators,
{\it albeit}, at extremely low temperatures. Based on first-principles density functional calculations, here we predict
that layered rhodium oxide K$_{0.5}$RhO$_2$ in noncoplanar chiral antiferromagnetic state is an unconventional
three-dimensional QAH insulator with a large band gap and a Neel temperature of a few tens Kelvins.
Furthermore, this unconventional QAH phase is revealed to be the exotic
quantum topological Hall effect caused by nonzero scalar
spin chirality due to the topological spin structure in the system and without the need of net magnetization and SOC.

\end{abstract}

\maketitle

%\section*{INTRODUCTION }

{\it Introduction.} The integer quantum Hall effect (IQHE), first found in 1980~\cite{klitzing},
is one of the most important discoveries in condensed matter physics.
When a strong perpendicular magnetic field is applied to a two-dimensional
(2D) electron gas at low temperatures, the Hall conductance is precisely quantized
in units of the fundamental conductance quantum (e$^2$/h) due to Landau-level quantization.
This quantization is subsequently found to be directly connected with
the topological property of the 2D bulk insulating states,
characterized by a topological invariant called the Chern number~\cite{tknn,Lau83}.
This topological understanding of the IQHE implies that the IQHE can also occur
in other time-reversal symmetry (TRS) broken systems with a topological non-trivial band structure
in the absence of the external magnetic field and Landau levels,
such as a ferromagnetic insulator, leading to the so-called quantum anomalous Hall effect (QAHE).
This effect was first proposed by Haldane in a honeycomb lattice model with a staggered magnetic field
that produces zero average flux per unit cell~\cite{haldane}.
Other model systems have also been proposed including the ferromagnetic
quantum wells in the insulating state~\cite{Liu08}, the disorder-induced Anderson insulator~\cite{Ono03},
Rashba graphene coupled with exchange field~\cite{Qia10,Che11}, Kagome lattice~\cite{Ogh10} and
ferromagnetic skyrmion crystal~\cite{Ham15}.

Due to its intriguing nontrivial topological properties and great potential application
for designing dissipationless electronics and spintronics, extensive theoretical studies
have been made to search for real materials to host such QAHE. The conventional mechanism for the QAHE
is the recognition of the QAHE as the quantized version of the anomalous Hall effect (AHE)
in a ferromagnetic metal~\cite{Nag10}. In particular, it has been recently established
that the Berry curvature in the momentum space caused by the broken TRS due to the magnetization and spin-orbit
coupling (SOC), acts as a fictitious magnetic field\cite{Fan03,Xia10} and thus gives rise to the AHE.
In a topologically nontrivial ferromagnetic insulator (Chern insulator), the integral of
the Berry curvature over the Brillouin zone (the Berry phase) results in a nonzero Chern number
and hence the quantized Hall conductance.\cite{weng} Indeed, several {\it ferromagnetic} insulators are
predicted to be the Chern insulators by first-principles band structure calculations
based on this mechanism\cite{ryu,weng}. Importantly, based on the prediction in \cite{ryu},
the QAHE has recently been observed in the Cr-doped (Bi,Sb)$_2$Te$_3$ ferromagnetic topological
insulator films\cite{xue}. Nevertheless, the QAH phase appears at extremely low tempetures
(less than 30 mK) due to the small band gap (less than 0.01 eV), weak magnetic coupling
and low carrier mobility in the sample. This hinders further
exploration of the exotic properties of the Chern insulator and also its applications.
The low carrier mobility could result from the disorder of the doped magnetic impurities in the sample.
Therefore, it would be fruitful to search for the QAHE in crystalline magnetic insulators with
a large band gap.

On the other hand, it was found in \cite{Ogh10} that in the Kagome lattice, the above-mentioned fictitious
magnetic field that gives rise to the AHE, can also be generated by the scalar spin chirality
$\kappa= \sum \vec{S}_i \cdot (\vec{S}_j \times \vec{S}_k)$ (where $\vec{S}_i, \vec{S}_j$  and $\vec{S}_k$ denote
three noncoplanar spins in the set)  due to the topological nontrivial spin texture
in the noncollinear magnetic structure. In such noncoplanar magnetic structure,
when an electron moves around a set of three noncoplanar magnetic moments, its wave function would acquire
a Berry phase of $\Omega /2$ where $\Omega$ is the solid angle subtended by
the three magnetic moments. This Berry phase acts as the fictitious magnetic field and generates the AHE
even in an antiferromagnet (AFM) without net magnetization.
Indeed, this unconventional AHE known as the topological Hall effect (THE) was
observed in noncollinear AFMs such as Nd$_2$Mo$_2$O$_7$ ~\cite{taguchi} and
Pr$_2$Ir$_2$O$_7$ ~\cite{machida2}.
Nevertheless, the topological Hall conductivity detected in these materials
are small, being only a small fraction of the conductance quantum (e$^2$/h)~\cite{taguchi,machida2,Sur14}.

Here we predict that the easily synthesized layered oxide K$_{0.5}$RhO$_2$ \cite{shibasaki,shyao}
in the noncoplanar antiferromagnetic
state (nc-AFM) (Fig. 1) would host the QAH phase with a large band gap of 0.22 eV,
based on a systematic first-principles study of its magnetic and electronic properties.
We also find that the QAH effect is caused by the nonzero scalar spin chirality in the noncoplanar AFM even in the absence
of the SOC and net magnetization, and thus is the quantized THE (QTHE). The calculated exchange coupling parameters
between the neighboring Rh atoms reveal that the nc-AFM is caused by the frustrated magnetic interactions
in the compound, with an estimated Neel temperature of $\sim$ 20 K.
All these findings suggest that the layered K$_{0.5}$RhO$_2$ is a promising candidate of the 
topological Chern insulator.

%\section*{ METHODS }
{\it Methods.} The electronic structure of K$_{0.5}$RhO$_2$ has been calculated based on the density functional
theory (DFT) with the generalized gradient approximation (GGA) plus on-site Coulomb repulsion 
(i.e., the GGA+U scheme) (see ~\cite{SM} for computational details).
The experimental lattice constants~\cite{yubuta} are used. Nevertheless, the
atomic positions are optimized theoretically.~\cite{SM}
Anomalous Hall conductivity (AHC) is calculated by using Wannier interpolation
with an effective Hamiltonian constructed in a basis of maximally localized Wannier functions (MLWFs)~\cite{wannier90}.
The band structure obtained from the effective Hamiltonian agrees well with that from
the DFT calculations (see Fig. S3 in ~\cite{SM}).

%\section*{ RESULTS }

\begin{figure}
\includegraphics[width=0.40\textwidth]{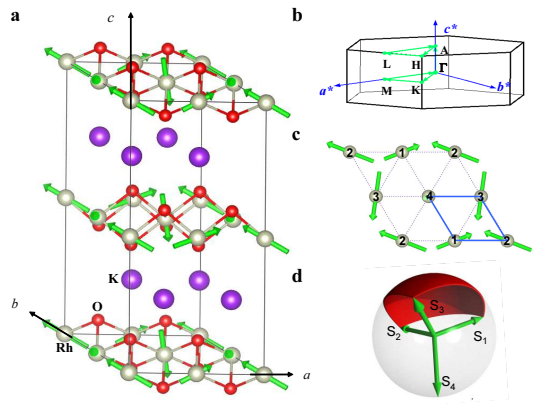}
\caption{\label{fig:str} (Color online) ({\bf a}) Crystal stucture of K$_{0.5}$RhO$_2$ in the $2\times 2 \times 1$ supercell
and ({\bf b}) its hexagonal Brillouin zone. 
Green arrows show the Rh magnetic moments in the ground state noncoplanar antiferromagnetic configuration (nc-AFM).
({\bf c}) Top view of the magnetic structure
in one RhO$_2$ layer. Thin blue lines denote the chemical unit cell. Numbers label the four Rh atoms in the
planar supercell. ({\bf d}) Unit sphere spanned by the four Rh magnetic moments
in one RhO$_2$ layer parallel transported to have a common origin.
}
\end{figure}

{\it Noncoplanar antiferromagnetic structure.}
Layered rhodium oxide  K$_{x}$RhO$_2$ has recently received increasing
attention~\cite{shibasaki,okazaki,shyao,saeed,zbb2} 
because it is isostructural and isoelectronic to thermoelectric material Na$_{x}$CoO$_2$~\cite{Sch03}.
K$_{x}$RhO$_2$ crystallizes in the $\gamma$-Na$_{x}$CoO$_2$-type structure 
where the CdI$_2$-type RhO$_2$ layer and the K layer stack alternately along $c$-axis\cite{shibasaki,shyao}, as illustrated in Fig. 1a.
Indeed, significant thermopower and Seebeck coefficient were observed in K$_{0.49}$RhO$_2$ ~\cite{shibasaki} and K$_{0.63}$RhO$_2$~\cite{shyao}, respectively.
Furthermore, as for Na$_{x}$CoO$_2$, K$_{x}$RhO$_2$ could be expected to become superconducting and also exhibit interesting
magnetic behaviors at certain potassium concentration ($x$), which, interestingly, can be tuned by K de-intercalation of KRhO$_2$~\cite{zbb2}.
In particular, Rh atoms in each RhO$_2$ layer form a two-dimensional triangular lattice which was recently shown to host
exotic magnetic states tunable by the band filling factor~\cite{martin,akagi}.
In  K$_x$RhO$_2$, Rh ions would be Rh$^{3+}$ (4$d^6$) when $x = 1$ and Rh$^{4+}$ (4$d^5$) if $x = 0$.
Therefore, Rh atoms in each RhO$_2$ layer would have their 4$d$ orbitals split into partially filled
$t_{2g}$ orbitals and empty $e_g$ orbitals~\cite{okazaki}, and $t_{2g}$ orbitals would be further split into
fully occupied double $e'_g$ orbitals and partially occupied single $a_{1g}$ orbital due to the trigonal deformation of RhO$_2$ octahedra.
Interestingly, when $x = 0.5$,  Rh ions would be Rh$^{3.5+}$ ($4d^{5.5}$) and the $a_{1g}$ band would have a filling factor of 3/4, which
was predicted by the mean-field solution of the ferromagnetic Kondo lattice model to have a chiral magnetic
ordering and spontaneous quantum Hall effect~\cite{martin,akagi}.

Here we investigate the magnetic properties of K$_{0.5}$RhO$_2$ with first-principles
DFT calculations.  We consider all possible magnetic configurations up to four-sublattice
orders on one RhO$_2$ monolayer (see Figs. S1 and S2 in ~\cite{SM}).
We perform total energy calculations 
for these magnetic configurations within the GGA+U scheme.
The calculated total energy and properties of the magnetic structures that could be stabilized
during the self-consistent calculations are listed in Table S1 in ~\cite{SM}.
It is clear from Table S1 that the all-in/all-out noncollinear nc-AFM 
configuration (Fig. 1) has the lowest total energy.
Furthermore, the nc-AFM configuration is an insulator with a band gap of 0.22 eV,
while all the other configurations are metallic.

\begin{figure}
\includegraphics[width=0.40\textwidth]{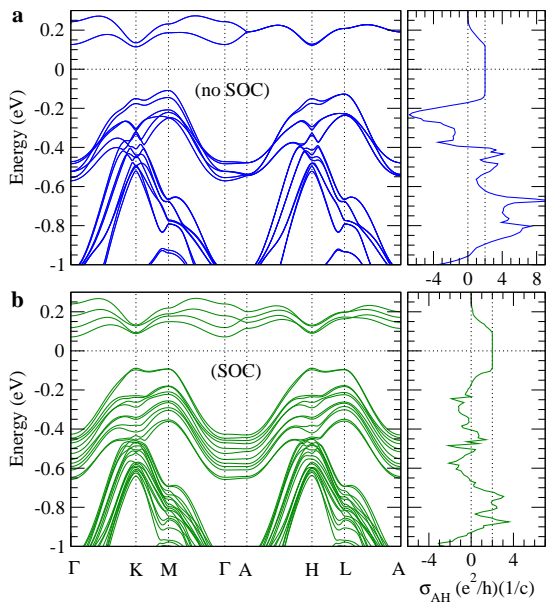}
\caption{\label{fig:band} Band structure and anomalous Hall conductivity ($\sigma_{AH}$)
of K$_{0.5}$RhO$_2$ in the noncoplanar antiferromagnetic state (nc-AFM)
without SOC ({\bf a}) and with SOC ({\bf b}).}
\end{figure}

{\it Quantum topological Hall insulating phase.}
Plotted in Fig. 2a is the band structure of K$_{0.5}$RhO$_2$
in the nc-AFM configuration. 
Figure 2a shows that the system is an insulator with a band gap of 0.22 eV.
To examine whether the band gap is topologically trivial or not, we further calculate
the AHC. 
For three-dimensional (3D) quantum Hall insulators, AHC $\sigma_{AH} = n$ e$^2$/h$c$ where
$c$ is the lattice constant along $c$-axis, $n$ should be an integer and equal to the Chern number ($n_C$) ~\cite{Hal87}.
The calculated AHC is displayed as a function of the Fermi level in Fig. 2a.
Indeed, we find that the band gap is topologically nontrivial with
the calculated Chern number $n_C = 2.0$, and hence the system is a 3D Chern insulator.
Figure 2b shows that the calculated AHC remains constant and equals to 2.0 e$^2$/h$c$ in the entire
band gap region.

Since the total magnetization of  K$_{0.5}$RhO$_2$ in the nc-AFM state is zero and
the SOC is not included in the electronic structure calculation yet, the obtained nonzero
AHC is not caused by spontaneous occurrance of magnetization and SOC\cite{Nag10}, and hence is
unconventional. Instead, the nonzero AHC results from the nonzero scalar spin chirality
$\kappa= \sum \vec{S}_i \cdot (\vec{S}_j \times \vec{S}_k)$
generated by the {\it noncoplanar chiral} magnetism in the nc-AFM state.\cite{Ogh10,taguchi}
In the nc-AFM magnetic configuration, there are four magnetic moments on
four Rh atoms in one RhO$_2$ layer (their unit vectors labelled as $S_1, S_2, S_3, S_4$ in Fig. 1).
By parallel transporting these four unit vectors to have a common origin, we obtain
an unit sphere, as shown in Fig. 1d. Therefore, the sum of the four solid angles ($\Omega$) spanned
by the four three-spin sets of $\vec{S}_1 \cdot (\vec{S}_2 \times \vec{S}_3)$, $\vec{S}_2 \cdot (\vec{S}_3 \times \vec{S}_4)$,
$\vec{S}_3 \cdot (\vec{S}_4 \times \vec{S}_1)$, $\vec{S}_4 \cdot (\vec{S}_1 \times \vec{S}_2)$
is 4$\pi$. The associated Berry phase $\gamma$ is then half of the total solid angle $\Omega$, i.e., $\gamma = \Omega/2 = 2\pi$,
and this gives rise to a Chern number of $n_C = \gamma /2\pi =1$.\cite{weng} Since one unit cell contains two RhO$_2$ layers,
the Chern number of the system would be 2 and the AHC $\sigma_{AH} = 2.0$  e$^2$/h$c$.
Therefore, the QAH insulating phase predicted here is entirely due to the QTHE. %quantized topological Hall effect.
Finally, we note that the sign of the Chern number and AHC could be reversed by reversing all the four
magnetic moments, e.g., changing the all-in to all-out nc-AFM configuration (see Fig. S2 in ~\cite{SM}).

\begin{figure}
\includegraphics[width=0.45\textwidth]{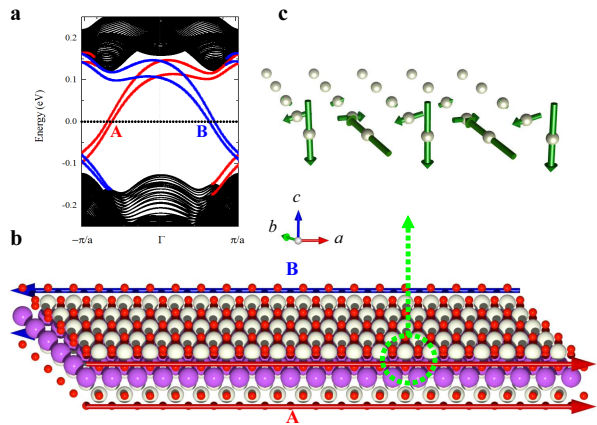}
\caption{\label{fig:wannier} ({\bf a}) Band structure of a thick noncoplanar antiferromagnetic K$_{0.5}$RhO$_2$
ribbon  cut along $a$-axis.  Black lines represent projected bulk energy bands,
while red (A) and blue (B) lines denote chiral edge states in the bulk band gap on the two opposite
sides of the ribbon.  In ({\bf b}),  red and blue arrowed lines represent the chiral edge states
A and B, respectively, one per RhO$_2$ layer.
({\bf c}) Spin texture of the A edge state near the Ferm level in a segment of the K$_{0.5}$RhO$_2$ ribbon.
Green vectores represent the spin moments on the Rh atoms.
}
\end{figure}

{\it Chiral edge states and spin texture.}
According to the bulk-boundary correspondence\cite{Ess11}, the nc-AFM K$_{0.5}$RhO$_2$ would have
one metallic chiral edge state per RhO$_2$ layer which carries dissipationless charge current.
To examine these interesting edge states, we calculate the energy bands for a thick K$_{0.5}$RhO$_2$
nanoribbon along $a$-axis using the MLWFs.
The thick nanoribbon has a width of 40 unit cells along $b$-axis and remains periodic along $a$- and $c$-axis.
The calculated one-dimensional band structure along $a$-axis is shown in Fig. 3a.
We can see that there are four edge states within the bulk band gap, two localized on one edge
carrying electrons along positive $a$ direction (red lines) and two sitting on the opposite edge
carrying electrons along negative $a$ direction (blue lines). In other words, there are two
chiral edge states per unit cell, i.e., one chiral edge state per RhO$_2$ layer,
as illustrated in Fig. 3b.
We note that the direction of the dissipationless edge current can be switched by changing the sign of
the scalar spin chirality, or by reversing, e.g., the all-in nc-AFM configuration to all-out one.

The spin texture of the edge state in K$_{0.5}$RhO$_2$ is studied by the one-band
tight-binding (TB) Hubbard model analysis (see ~\cite{SM} for details).
Figure 3c displays the spin texture of the A edge state near the Fermi level in a segment of the K$_{0.5}$RhO$_2$
ribbon. We find that the spin moments are well localized on the atoms along the edges and decay rapidly
towards the interior of the ribbon.

%\section*{DISCUSSION}

%{\it Effect of spin-orbit coupling.}
{\it Discussion.} In the above-mentioned electronic structure calculations, the SOC was not included yet.
Therefore, the uncovered quantized AHC can be completely attributed to nonzero scalar
spin chirality, 
and hence results from the genuine QTHE. Nevertheless, the SOC always exists
in real materials. In particular, the SOC strength of Rh 4$d$ orbitals is not small and
hence cannot be ignored. To examine how the electronic structure and especially QTHE
discovered here would be affected by the SOC effect, we  have repeated the calculations
with the SOC included, and the calculated energy bands and AHC are shown in Fig. 2b.
It is clear from Fig. 2b that almost all doubly degenerate bands are now split due to
the SOC. Figure 2b also indicates that the size of the calculated AHC below the band gap
generally gets reduced when compared to Fig. 2a, although the line shape remains similar.
Importantly, Fig. 2b shows that the system is still an insulator with a similar band gap
(0.16 eV), and also has a quantized AHC with the Chern
number of 2. Interestingly, when the SOC is included, the system acquires a small
magnetization of 0.08 $\mu_{\mathrm{B}}$/f.u. due to the spin canting. This small magnetization
would provide a coupling between the magnetic structure and external magnetic field
and thus allows us to control the sign of the Chern number and the direction of
the edge currents by an applied magnetic field.

%{\it Exchange coupling parameters.}
In order to understand the noncollinear nc-AFM order and also estimate its Neel temperature,
we evaluate the exchange coupling parameters between neighboring Rh atoms in one RhO$_2$ layer
by mapping the calculated total energies of the ferromagnetic (FM), striped (s-AFM)
and zigzag (z-AFM) antiferromagnetic states (see Table S1 in \cite{SM}) to the classical Heisenberg
model.
In so doing, we obtain the first near-neighbor exchange coupling $J_1= 4.4$ meV =  51 K (FM coupling)
and second near-neighbor exchange coupling $J_2 = -3.6 $ meV = -42 K (AFM coupling).
According to the phase diagram of the $J_1$-$J_2$ Heisenberg model~\cite{Saa15},
this implies that noncollinear magnetic states are energetically favored in K$_{0.5}$RhO$_2$.
Furthermore, based on these parameters, a mean-field estimation (see Ref. \cite{Tun11} and references therein)
would lead to a Neel temperature of $\sim$20 K for K$_{0.5}$RhO$_2$, being well above 30 mK
at which the QAH effect was observed in thin films of Cr-doped (Bi,Sb)$_2$Te$_3$, a ferromagnetic
topological insulator~\cite{xue}. Thus, the QTHE predicted in K$_{0.5}$RhO$_2$ here
would provide an essily accessible platform for exploring exotic states of quantum matters
and also be promising for technological applications.

\begin{figure}
\includegraphics[width=0.45\textwidth]{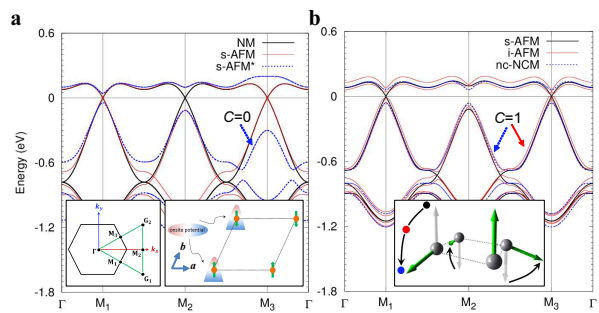}
\caption{\label{fig:TB-bands} ({\bf a}) The TB energy bands of the NM,
s-AFM and s-AFM* along the symmetry lines in the Brillouin zone (left inset).
The s-AFM* state differs from the s-AFM in that an onsite potential is added
on one of the two sites on each ferromagnetic (FM) chain along $a$-axis (right inset).
({\bf b}) The TB band structures of the s-AFM and nc-AFM configurations as well as
an intermediate noncollinear magnetic state (i-NCM)
(inset) which occurs when the s-AFM is transformed to the nc-AFM by spin rotations.
}
\end{figure}

%{\it Tight-binding model analysis.}
To gain detailed insight into the formation of topological insulating gap,
we also perform a TB Hamiltonian analysis of the magnetic and electronic
properties of one monolayer of K$_{0.5}$RhO$_2$. We consider
the one-band TB Hubbard model because the two conduction bands and six top valence
bands of K$_{0.5}$RhO$_2$ in the nc-AFM structure (Fig. 2a) are derived mainly from
the Rh 4$d$ $a_{1g}$ orbital. We solve the model Hamiltonian self-consistently
within the mean-field approximation (see ~\cite{SM} for details).
We find that the nc-AFM magnetic structure (Fig. 1)
is the most stable state, being consistent with the DFT calculations.
Figure 4a shows that the TB band structure of the NM state 
is a metal with conduction and valence bands touching at all three M-points in the BZ at the Fermi level.
In the primitive unit cell, the Rh $a_{1g}$ band is 3/4-filled and hence metallic.
Therefore, in the $2\times2$ supercell, three of the four $a_{1g}$ bands obtained
by folding the energy bands in the primitive unit cell, are completely filled up
to the BZ boundary including the M-points. However, the supercell posseses
three half-translation symmetries ${\mathbf{a}}/{2}$, ${\mathbf{b}}/{2}$ or ${\mathbf{(a+b)}}/{2}$,
and consequently, the conduction and valence bands touch and the bands are doubly
degenerate at the M-points.
Introducing the striped AFM (s-AFM) (Fig. S1b)
configuration breaks the ${\mathbf{b}}/{2}$ and ${\mathbf{(a+b)}}/{2}$ half-translation symmetries,
although preserves the half-translation ${\mathbf{a}}/{2}$,
and thus opens a gap at the M$_2$-point in the BZ. Putting an extra onsite potential
on two sites in the supercell (see the right inset in Fig. 4a)
further lifts the ${\mathbf{a}}/{2}$ half-translation symmetry and opens a gap at the M$_1$ and M$_3$ points.
Therefore, the resultant AFM* state is an insulator. Nevertheless, it is not a QAH
insulator because the calculated Chern number $n_C=0$.

Similarly, the nc-AFM configuration also breaks the three half-translation
symmetries and thus opens a gap at all three M-points (Fig. 4b). In contrast, however,
the gap is nontrivial because the calculated $n_C=1$. Note that the nc-AFM
configuration can be obtained from the s-AFM state by rotating the spins on
the four sublattices (see the inset in Fig. 4b).
Interestingly, all these intermediate
states are a QAH insulating state with $n_C=1$, except the s-AFM state where the Chern
number is ill-defined because it is a metal. This is caused by the fact that
all the intermediate states have a nonzero scalar spin chirality with total solid
angle spanned by the four spins is 4$\pi$, as for the nc-AFM state.
These results clearly demonstrate that the QAH phase found here is robust against
variation of the magnetic structure around the nc-AFM.

%{\it Quantum topological Hall effect in ferromagnets.}
As mentioned before, the nc-AFM order was recently predicted in the ferromagnetic Kondo triangular lattice model
due to the perfect Fermi surface nesting at the 3/4 band filling~\cite{martin,akagi}.
As found here, the nc-AFM phase is insulating and exhibits spontaneous quantum Hall effect\cite{martin,akagi}.
Thus, given the well known predictive power of the DFT calculations, the QAH phase discovered
here in K$_{0.5}$RhO$_2$ is a materialization
of the prediction of the QTHE in the ferromagnetic Kondo triangular lattice model~\cite{martin,akagi},
{\it albeit}, with a distinct origin.
We also notice that Hamamote \textit{et al.} very recently predicted the QTHE in the skyrmion crystal
based on the double-exchange model calculations~\cite{Ham15}.
In this model, the conduction electrons are assumed to strongly couple to the background skyrmion spin
texture and thus acquire a Berry phase when they hop among the noncollinear spin moments.
Therefore, the mechanism is the same as that in the ferromagnetic Kondo lattice model
in which the noncollinear local spin moments of the magnetic ions, instead of the skymions,
provide the fictitious magnetic field and Berry phase. However, the QTHE in the skyrmion
crystal ~\cite{Ham15} differs in several significant ways from the one predicted in the present paper.
For example, a strong SOC (e.g., Dzyaloshinskii-Moriya interaction or
Rashba interaction) is needed to stabilize the skyrmion crystal~\cite{Ham15}.
Furthermore, a skyrmion crystal is formed in a ferromagnet.

%\begin{acknowledgments}
This work is supported by the State Key Program for Basic Research (2015CB659400 and 2013CB632700),
the National Science Foundation of China (Nos. 11474150, 11574215, 11274359, and 11422428) and the Scientific
Research Foundation for the Returned Overseas Chinese Scholars, Ministry of Education of China
 as well as the Ministry of Science and Technology, National Center for Theoretical Sciences
 and Academia Sinica in Taiwan.
J.Z. and Q.F.L. contributed equally to this work.
%\end{acknowledgments}

\end{document}